\input stromlo

\title Are ``dwarf'' ellipticals genuine ellipticals?
 
\shorttitle Are ``dwarf'' ellipticals genuine ellipticals?
 
\author Helmut Jerjen^1 and Bruno Binggeli^2
 
\shortauthor Jerjen \& Binggeli
 
\affil  @1 Mount Stromlo and Siding Spring Observatories, ANU, Australia
 
\affil  @2 Astronomical Institute of the University of Basel, Switzerland
 
\abstract We review the systematic properties of ``dwarf'' elliptical (dE) 
galaxies, focussing on the relation between ``normal'' and ``dwarf'' 
ellipticals. In recent years, this relation has been described as ``dichotomy'' 
-- based essentially on a discontinuity in central surface brightness. We 
show that, outside of 300 pc from the centre, the S\'ersic profile parameters 
vary continuously from ``normal'' to ``dwarf'' ellipticals. The ``dichotomy'' 
is indeed restricted to the very central part, where differences also exist 
among ``normal'' ellipticals (E). Bright, nucleated dE's closely resemble 
``normal'' E's also in their clustering and flattening properties. They may 
be genuine ellipticals, having no present-day late-type dwarf precursor from 
which they could have been manufactured. The non-nucleated dE's ("dwarf spheroidals") 
may be a different breed.        
 
\section What to call a dE galaxy? 
 
Elliptical galaxies are distinguished from spirals and late-type dwarfs by their 
smooth light distribution. Fainter than $M_{B_T}\sim-18$ they divide up into two classes: 
compact ellipticals with high surface brightness, exemplified by \index{NGC221} M32, and diffuse 
ellipticals with low surface brightness, exemplified by the dwarf spheroidals in the Local 
Group (LG). Many different terms like \index{dwarf ellipticals} dwarf ellipticals, dwarf spheroidals, 
spheroidals, or LSB galaxies are in use for the second class; there is no generally accepted 
definition. This led to some confusion. In particular, one debates whether faint ellipticals 
like M32 should be called dE. In our discussion we adopt the classification scheme worked 
out and illustrated in the Virgo cluster dwarf atlas of Sandage \& Binggeli (1984),  
where the term ``dwarf ellipticals'' encompasses both local dwarf spheroidals and similar 
looking galaxies beyond the LG. Faint ellipticals with {\em high} surface brightness are 
referred to as ellipticals or \index{compact ellipticals} compact ellipticals (cE) but never 
dwarf ellipticals. For a detailed discussion of the nomenclature issue see Binggeli (1994) 
and Kormendy \& Bender (1994).
 
Within the dwarf elliptical family there is a subtype classified as \index{dwarf lenticular} 
dwarf S0 (dS0). These galaxies are among the brightest dwarf ellipticals, being a very rare species. In the Virgo cluster, 
for instance, only 25 dS0's are known as compared to 800 dE's. The dS0's show morphological 
characteristics of a bulge-to-disk transition which is typical for classical S0's. 
However, a bar feature or simply high apparent flattening (Sandage \& Binggeli 1984, panel 8) are 
other reasons why a dwarf galaxy is called a dS0 rather than a dE. Interestingly, dS0's seem to 
be statistically indistinguishable from bright dE's 
with respect to their mean radial surface-brightness profile (Binggeli \& Cameron 1991 
hereafter BC91) and other photometric parameters (Ryden et al.~1997). However, first 
kinematical data for a small number of dE\&dS0's give preliminary evidence that galaxies 
tagged with the label ``dS0'' are the only rotationally supported early-type dwarfs (Bender 1997). 
 
\section The E--dE dichotomy and how it disappears
 
One of the classical representations of the galaxian manifold is a plot of the 
absolute magnitude $M$ versus the observed central surface brightness 
$\mu_0$ (Kormendy 1985; Binggeli 1994). In that plane, the two elliptical 
families appear to fall into two distinct sequences. While E's and E-like bulges 
have higher central surface brightness with fainter luminosity, these parameters are 
anti-correlated for the dE's. Until recently there was the hope to get resolved core 
photometry for more distant low-luminosity ellipticals in the magnitude range 
$-18<M_{B_T}<-15$, where members of both E's and dE's coexist, to answer the question 
whether there is a bridge between the two families or a gap. However, Kormendy and collaborators 
(Kormendy et al.~1994; Kormendy \& Bender 1994) showed convincingly that even 
with the power of CFHT and HST the cores of these galaxies remain unresolved down to a radius 
of $0.1\,$arcsec. It will thus be impossible in the near future to discern where these key 
objects are located in the $M$--$\mu_0$ diagram.  A presently more promising approach  
to the problem of the \index{dichotomy, E--dE} E--dE dichotomy is to explore the deviations 
of light profiles from the classical $R^{1/4}$-law or \index{exponential law} 
exponential law, which we discuss in the following.
 
In first order the surface brightness profiles of dE's are well approximated 
by straight lines reflecting the exponential decay of the light intensity with 
radius (Faber \& Lin 1983). But it was emphasized by Caldwell \& Bothun (1987) and BC91  
that there are {\em systematic} deviations from this exponential law in the 
central region. Most dE's brighter than $M_{B_T}=-16$ have an inner luminosity excess 
above the exponential (BC91, Fig.~8), which is {\em not} due to the star-like nucleus 
some of these dwarfs have, but to a shallow extension over several hundred of 
parsecs. On the other hand, very faint dwarfs typically exhibit a central decrement 
relative to an exponential law. A closer inspection of a collection of Virgo dwarf profiles 
(BC91, Fig.~4) reveals that the \index{profile shape} {\em shape}, or {\em curvature} of the
profile varies with total luminosity. The link between the two elliptical families are 
the brightest dE's, which come confusingly close to normal E's in their profile shape.
The observed variation motivated Young \& Currie (1994, actually preceded by Davies et 
al.~1988) to approximate dwarf profiles with the generalized exponential function 
initially introduced by \index{S\'ersic profile} S\'ersic (1968): $I(r)=I_0\exp[-(r/r_0)^n]$, 
$0<n$, where $I(r)$ is the light intensity at radius $r$, $I_0=I(0)$, and $r_0$ is the scale length. 
Note that most of our ``giant colleagues" are using $1/n$ instead of $n$. The 
corresponding integrated light profile (growth curve) is given by $GC(r)={2\pi I_0 r_0^2 \over n} 
\cdot\gamma[2/n,(r/r_0)^{n}]$ where $\gamma(a,x)=\int_0^{x}\exp(-s)s^{a-1}ds$.

Obviously, the S\'ersic profiles encompass both the $R^{1/4}$-law and the 
exponential law ($n=1$) by allowing the shape parameter $n$ to vary. 
For $n>1$ the profiles become flat in the central part, just as observed for 
faint dE's (see Fig.~1). What was qualitatively known from Virgo dwarf profiles 
was thus quantified by Young \& Currie (1994), who found a strong correlation between 
$n$ and total luminosity for a sample of Fornax cluster dwarfs. The authors even 
used the relation to measure the distance to the Fornax cluster and later applied 
it also to the Virgo cluster (Young \& Currie 1995), where the observed scatter 
in the $n$--$M$ relation is much larger. Young \& Currie ascribe the large scatter 
to the depth of the Virgo cluster, assuming a universally small dispersion of 
the $n$--$M$ relation. However, Binggeli \& Jerjen (1996) show that this assumption 
is flawed and that the claimed filamentary structure of Virgo is unreal. The 
$n$--$M$ relation is probably of no use for \index{distance indicator} distance 
measurements. For details the reader is referred to Binggeli \& Jerjen (1996).
 
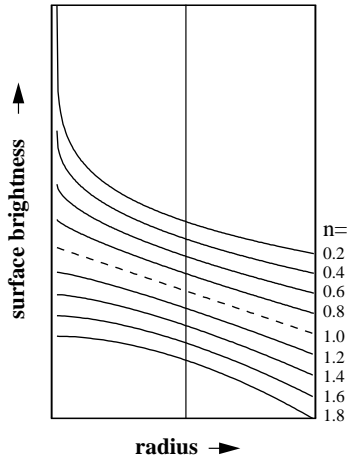
\figureps[JB97afig1.epsf,0.4\hsize] 1. Curvature variation of the S\'ersic 
profiles as a function of the shape parameter $n$. The vertical line indicates the 
radius at which all functions have the same slope.
 
To investigate the shape variation in greater detail we have analysed the 
light profiles of an unbiased sample of dE\&dS0's from the photometric survey of Virgo 
dwarfs (BC91; Binggeli \& Cameron 1993). The galaxy sample is complete down to 
$M_{B_T}=-14$. The data are based on high-resolution photographic plates from the 
Las Campanas 100-inch du Pont telescope. We fitted the $GC(r)$ function as given 
above to the growth curve of each sample galaxy. Errors in the intensity counts 
were assumed to be Poissonian. Because our prime interest is in the {\em universal} 
shape signature of the profile, we spared out the innermost $3''$ of the profiles 
(or $\sim300\,$pc, assuming a distance modulus of 31.75 for Virgo, Sandage \& Tammann 1995). In 
this way central features such as the nuclei some of these dwarfs have are excluded 
from the fit. The outer fit limit was taken at the surface brightness radius 
$r_{27}$. In Fig.~2 we plot $\log(n)$ versus luminosity for our Virgo sample (open 
symbols). The dwarfs follow a clear trend from larger $n$ for faint dwarfs to smaller 
values for brighter galaxies, as it was reported in earlier studies. However, the scatter
of the relation is considerable with $\sigma_{\log(n)}=0.14$ and 
$\sigma_{M_{B_T}}=1.37$. This indicates the large error which has to be expected 
from distances based on this method. V615 appears to be the only 
dE not following the general trend. The reason is its very extended nucleus ($r_{nuc}>3''$, 
Fig.~4 in BC91) which affected the profile fit in this particular case. Thus the galaxy 
can not be seen as a real deserter. We compare our dwarf data with a parameter set published 
for E\&S0's by Caon et al.~(1993). Their good and fair quality fits have been 
added in our diagram as filled symbols. Obviously, they follow the same trend as the dwarfs, 
with about the same scatter. The \index{scaling law, $n$--$M$} relation for E's and that for dE's smoothly and 
continuously merge into each other, giving the impression of one global relation for dwarf and
giant ellipticals over an $8\,$magnitude range. 
A ML fit gives: $\log(n)=1.40(\pm0.10)+0.10(\pm0.01)M_{B_T}$.
Even the supergiant cD galaxies seem to fit into this sequence of profile shapes (Graham et al.~1996).
 
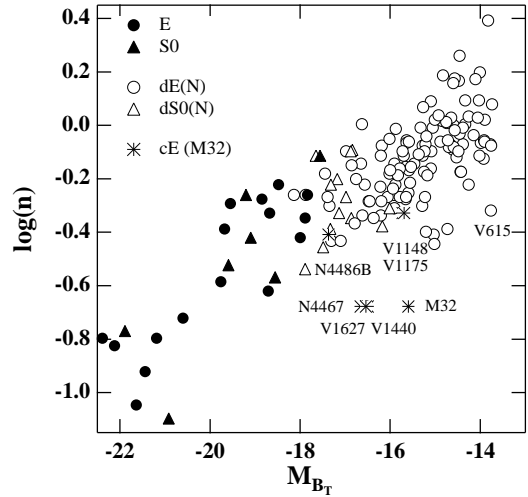
\figureps[JB97afig2.epsf,0.6\hsize] 2. The $\log(n)$--luminosity relation for 
early-type galaxies. 
 
In the diagram we have further added data for cE galaxies, which include about half of the 
known Virgo cE sample (\index{NGC4486B} N4486B, \index{NGC4467} N4467, \index{VCC1148} V1148, 
V1175, V1440, and V1627) plus \index{NGC221} M32, for which we have fitted a generalized profile to the 
photometric data of Peletier (1993). Within a narrow range of luminosity, this galaxy type shows a large variety of 
shapes, obviously not following the trend of the other ellipticals. An individual look 
with respect to projected distance and relative velocity to a bright parent 
galaxy gives some hint that the deviation from the ``main sequence'' 
might be correlated with the degree if isolation. V1627, 
V1440, and M32 are close companions of a giant galaxy (M89, N4548, and M31,
respectively), where tidal stripping during or after formation may have 
occurred (Faber 1973, Burkert \& Truran 1994). On the other hand, V1148 and V1175 
appear to be rather isolated in the Virgo cluster. Less clear is the situation for N4467 
and N4486B where relative velocities to possible parent galaxies are high 
(N4467: $\Delta v=501\,$km$\,$s$^{-1}$ with M49; N4486B: 
$\Delta v=228\,$km$\,$s$^{-1}$ with M89). The same is seen in \index{scaling law, $\mu_0$--$M$} 
Figs.~3 and 4 \index{scaling law, $\log(r_0)$--$M$} 
where we show the correlations between the two other S\'ersic parameters
$r_0$ and $\mu_0$ and luminosity. In both diagrams the cE's are the 
only type which deviates clearly from the relation exhibited by the dE\&dS0's 
and E's. Overall, these results support the view that the cE's are a special
kind of elliptical galaxies; they have shapes like giants but luminosities
like dwarfs. It is unlikely that they are low-luminosity representatives of
giant ellipticals. Bender et al.~(1992) suggested that cE galaxies may be bulges 
of failed disk galaxies that could not acquire a significant disk component due 
to the tidal field of a nearby massive galaxy. However, this possibility seems to  
contradict the observed $\log(n)$--luminosity relation for bulges of spiral 
galaxies (Andreakis et al.~1995). 
 
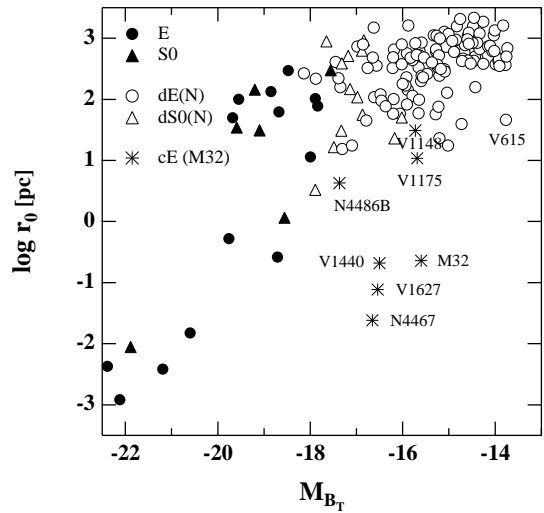
\figureps[JB97afig3.epsf,0.6\hsize] 3. The scale length--luminosity relation 
for early-type galaxies. 

Fig.~4 is of special interest because it represents the model-based analogy 
of the $M$--$\mu_0$ diagram mentioned in the beginning. Compared to the model-free 
($M$, $\mu_0$) values, the S\'ersic fits for brighter ellipticals ($M_{B_T}<-21$) yield 
central surface brightnesses which are systematically {\em higher}. This shift resolves the E--dE 
dichotomy by moving up all these galaxies onto the locus of the universal relation, 
i.e.~to brighter central surface brightness with increasing luminosity followed by all ellipticals
(except the cE's) below this magnitude limit. It is well-known that at about 
this luminosity ($M<-21$) many other properties within the E family are changing. It is roughly 
the transition point from resolved to unresolved cores (Kormendy et al.~1994); from 
boxy to disky shapes (Nieto et al.~1991); from E1.5 to E3 apparent flattening (Tremblay \& 
Merritt 1996); and from anisotropic to rotationally supported systems (Davies et al.~1983). From 
our profile shape correlation (Fig.~1) we would ``predict", by extrapolating inwards,
a high central surface brightness for the brightest ellipticals; which is not 
observed in the resolved cores. A brief discussion about the reason for 
this apparent discrepancy leads us first to the question about the origin 
of the $\log(n)$--luminosity scaling law. 

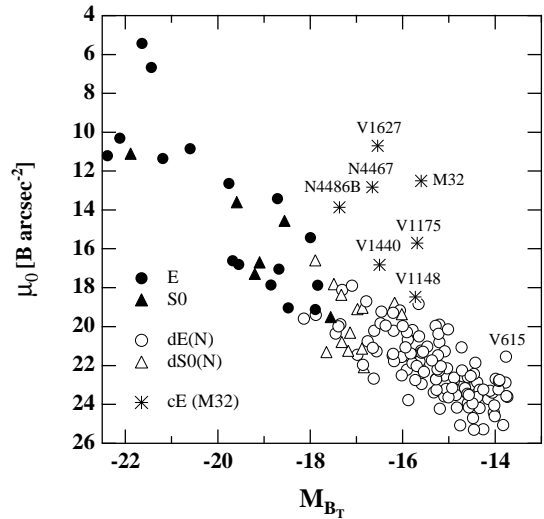
\figureps[JB97afig4.epsf,0.6\hsize] 3. The central surface brightness--luminosity 
relation for early-type galaxies. 

For the E\&S0 galaxies it was shown (Einasto \& Caon 1993) that $n$ is not correlated 
with the underlying environmental density. This result is consistent with the fact that 
dwarfs and giants have both very similar clustering properties but obviously different 
profile shapes. It must be rather something intrinsic, such as the total mass, which 
determines the global light distribution. If similar formation mechanisms were at work 
for these galaxies, the only difference might be the depth of the gravitational 
potential. This idea has been suggested by Young \& Currie (1994) and 
Andredakis et al.~(1995) and is getting some theoretical support from
models of violent relaxation processes (Hjorth \& Madsen 1995). In the case of
dwarf galaxies there is common agreement that the specific light distribution,
i.e.~the diffuseness, is a result of low mass. With decreasing mass (fainter 
luminosity) the potential becomes more shallow and stellar processes are shaping 
the appearance of the system accordingly. Supernova-driven winds and winds of massive 
stars remove the gas from the centre and subsequently star formation will shift to 
the outer regions of the galaxy leading to a plateau-like light profile. 
At higher luminosities the ellipticals have a more and more cuspy light 
distribution along with an extended halo (cf.~Fig.~1) -- both consequences of a 
small value of $n$. However, what is observed is a light deficiency relative to 
the overall profile shape concentrated on the innermost region ($r<300$pc) of 
bright ellipticals ($M_{B_T}<-21$). Here it is important to note that the core regions of 
elliptical galaxies are special and unique. There one finds black 
holes (Kormendy \& Richstone 1995) as well as stellar and gaseous disks 
(Kormendy et al.~1994; see also these proceedings). Hence, the {\em observed} $\mu_0$ 
of a giant elliptical is {\em not} a good tracer of its global galactic 
properties but is rather a foot print of the individual (dynamical and dissipational) 
history of its core region.

Outside the innermost $300\,$pc, normal and dwarf ellipticals show a great continuity 
in their structural properties. In contrast to the previous emphasis of a dichotomy 
between giants and dwarfs (Wirth \& Gallagher 1984, Kormendy 1985, BC91), 
this suggests that {\em the diffuse, low-surface brightness dwarf galaxies are the true 
low-luminosity extension of the classical giant ellipticals}. In the following we will 
summarize other properties common to both galaxy families to provide further evidence 
that in many respects they are indistinguishable. A review on the properties of dwarf 
elliptical galaxies in general can be found in Ferguson \& Binggeli (1994, hereafter FB94).    
 
\section More systematic properties in comparison

\subsection Flattenings and kinematics 
 
Binggeli \& Popescu (1995) studied 260 photometrically measured and 
\index{dwarf ellipticals, flattening}\index{dwarf ellipticals, kinematics}
800 eye-estimated apparent ellipticities of all different 
types of dwarf galaxies in the Virgo cluster. They found good agreement 
between the flattening distributions of giant ellipticals 
(Franx et al.~1991) and  {\em nucleated} dE's. Among the dE\&dS0 galaxies, the 
nucleated dE's tend to be significantly rounder than 
their non-nucleated counterparts. This was noted before by Ferguson \& Sandage 
(1989) and Ichikawa (1989) and confirmed the results 
by Ryden \& Terndrup (1994). Non-nucleated dE's, late spirals, Im's, and 
BCD's show all very similar distributions, with some 
hint that dE's and Im's are slightly rounder than the rest. The flattening 
distribution of the dS0's is comparable with that of 
spiral galaxies, indicating the disky nature of these systems. Concerning the 
3-dimensional shape, the situation for the early-type 
dwarfs seems to be quite similar to normal ellipticals. A modest triaxiality 
explains the ellipticity distribution of dE's much 
better than a pure oblate model. 

The question of rotation in dwarf ellipticals has been addressed only 
recently because of required spectroscopic work at a faint level of 
surface brightness. The sparse results for only six galaxies suggest that 
dwarf ellipticals are not supported by rotation (see FB94, Sect.~3). This excludes 
at least some of the dS0's, e.g.~UGC7436, a dS0,N in the 
Virgo cluster (Bender 1997). However, a much larger kinematic sample is needed to 
explore the kinematic nature of dwarf ellipticals and the possible differences 
between dE's and dS0's.  
 
\subsection Luminosity function and spatial distribution
 
\index{dwarf ellipticals, luminosity function} \index{dwarf ellipticals, spatial distribution}
The intermediate magnitude range between faint E's and faint dE\&dS0's is 
populated by the nucleated dE's and dS0's. These 
brightest dwarf ellipticals are unique in the sense that they can only be 
found in clusters or as close companions to massive 
parent galaxies. In fact, all known dE\&dS0's brighter than about 
$M_{B_T}-16$ are cluster members. For instance, the brightest 
dE\&dS0's in Virgo, Fornax, or Centaurus are -18 (Sandage et al.~1985), 
-17.5 (Ferguson \& Sandage 1988, hereafter FS88), and -18 (Jerjen \& 
Tammann 1996), respectively. On the other hand, N205 is the brightest dwarf 
elliptical in the LG with only $M_{B_T}=-15.6$, and the 
nearby Cen$\,$A group has no dwarf brighter than $M_{B_T}=-13$ (Jerjen et 
al.~1996). A trace of this special population of dwarf ellipticals 
in clusters can be found in the shape of the type-specific dE\&dS0 
luminosity function (LF). The LF of the Virgo dwarf ellipticals 
(Sandage et al.~1985, Fig.~6) exhibits a clear plateau with a maximum at 
$M_{B_T}\sim-15$ before it rises steeply to fainter 
magnitudes. A similar signature can be found in the LF of Fornax cluster dE\&dS0's 
(FS88, Fig.~17). One could think of a superposition of a 
Gaussian and a Schechter function representing the nucleated and non-nucleated 
dwarf populations, respectively.
 
It is a general result from the morphology-density relation of dwarfs 
(FS88; Binggeli et al.~1990; Vader \& 
Sandage 1991) that early-type dwarfs are the most strongly clustered of all 
galaxy types. However, there seem to exist isolated, faint, non-nucleated dE's 
such as the recently discovered Tucana dwarf, which indicates again the importance
of a discrimination between the different subtypes to get the full 
information. Binggeli et al.~(1987) have qualitatively shown that
the nucleated dE's follow the strongly clustered projected distribution of 
giant ellipticals in the Virgo cluster. The non-nucleated dE's 
are slightly more spread out. This trend gained further support from a 
distribution study by Ferguson \& Sandage (1989) 
where significant differences could be found between nucleated and 
non-nucleated dE's brighter than about $M_{B_T}=-14$. The latter
follow the shallow cluster profile exhibited by the spirals and irregulars, 
hinting at a possible evolutionary link between non-nucleated dE's 
and late-type galaxies. From a recent deep redshift survey of the Centaurus
cluster (Stein et al.~1996) we also got first results
from the third dimension (velocity). The bright dE\&dS0's are the only galaxy 
type which has (1) a Gaussian velocity distribution 
and (2) a mean redshift similar to the cluster mean, demonstrating that they 
trace the highest densities. All other galaxy types 
exhibit non-Gaussian, irregular velocity distributions. 
 
\subsection Are dE's evolved irregulars?

\index{dwarf galaxies, irregulars} \index{dwarf galaxies, evolution sequence} 
The evolutionary connection between dE\&dS0's and irregular dwarfs 
Im\&BCD's is not well understood. Although many intermediate-type 
dwarfs were studied where we may be witnessing the conversion of an irregular 
to a dE (e.g.~ESO359-G29, Sandage \& Fomalont 1993), 
and several physical processes are known which can serve as transformation 
mechanisms (FB94), no consistent picture of  
dwarf galaxy evolution has emerged yet. 
What has become clear over the last 10 years, though, is that it will be very
hard, if not impossible to manufacture a bright, cuspy dE from a bright
late-type dwarf like the LMC by {\em any}\/ mechanism. 
Simple gas removal by ram pressure stripping would leave the remnant with 
much too low surface brightness, including a missing nucleus, as well as too
low metallicity (Binggeli 1986, Bothun et al.~1986, Davies \& Phillipps 1988,
FB94).
Clearly, the gas would have to be turned into stars instead of being lost,
but it is not clear {\em how}. Also, there is a kinematic dichotomy between
the non-rotating and roundish dE's and the rotating, disky irregulars.
It seems that the present-day bright irregulars are not the precursors of 
the bright dE's. In contrast, there is no such problem for the faint dwarfs. 
Faint irregulars are thick and slowly rotating. They are easily 
turned
into dwarf spheroidals, some of which, like Carina, show indeed the sign of
fairly recent star formation (see FB94). Hence also with
respect to secular evolution, at least the {\em bright}\/ dwarf ellipticals 
($M>-16$) are as close or as far from late-type galaxies as normal
ellipticals.   
 
\section Conclusions
 
We have shown the existence of a {\em global} and {\em continuous} relation 
between S\'ersic's profile shape parameter $n$ and absolute magnitude for 
E and dE galaxies. The continuity in luminosity profile characteristics 
holds outside the innermost $300\,$pc of the galaxies. The E--dE dichotomy,
i.e. high-surface brightness of normal E's versus low-surface brightness of
dwarf E's at intermediate luminosities, is restricted to the core region, where
``special effects" (stellar disks, black holes, prolonged dissipation etc.)
begin to dominate over the global structure. We have listed further family
bonds between normal and dwarf ellipticals: both have very 
similar flattenings and clustering properties, and at least the bright dwarfs
cannot be explained as evolved irregulars. One weak point here may be
the kinematics: faint normal ellipticals are rotation-supported, bright dwarf
ellipticals are apparently anisotropic; but the data for dwarfs is still too
sparse for a final statement. Overall it appears that ``normal" and "dwarf"
ellipticals form {\em one}\/ sequence, {\em one}\/ family of stellar systems 
which must have a common origin. In this sense dwarf ellipticals are the genuine 
low-luminosity extension of giant ellipticals, and the question put in the 
title of this contribution has to be answered with an emphatic YES!

{\em We thank the Swiss National Science Foundation for financial support.} 

\references
 
Andreakis, Y.C., Peletier, R.F., Balcells, M., 1995, MNRAS, 275, 874  
 
Bender, R., 1997, private communication
 
Bender, R., Burstein, D., Faber, S.M., 1992, ApJ, 399, 462
 
Binggeli, B., 1986, in: {\em Star-Forming Dwarf Galaxies}, eds.~D.~Kunth 
et al., \'{E}ditions Fronti\`{e}res, Gif-sur-Yvette, p.~54 
 
Binggeli, B., 1994, in: ESO/OHP Workshop on Dwarf Galaxies. 
Meylan, G. \& Prugniel, P., eds., ESO, Garching, p.~13
 
Binggeli, B., Cameron, L.M., 1991, A\&A, 252, 27 (= BC91)
 
Binggeli, B., Cameron, L.M., 1993, A\&AS, 98, 297
 
Binggeli, B., Jerjen, H., 1996, in preparation
 
Binggeli, B., Sandage, A., Tammann, G.A., 1985, AJ, 90, 1681
 
Binggeli, B., Tarenghi, M., Sandage, A., 1990, A\&A, 228, 42
 
Bothun, G.D., Mould, J.R., Caldwell, N., McGillivray, H.T., 1986, AJ, 92, 1007 

Burkert, A., Truran, J.W., 1994, in: {\em Panchromatic View of Galaxies -- Their 
Evolutionary Puzzle}, eds.~Hensler et al., Editions Frontieres, Gif-sur-Yvette, p.~123

Caldwell, N., Bothun, G.D., 1987, AJ, 94, 1126
 
Caon, N., Capaccioli, M., D'Onofrio, M., 1993, MNRAS, 265, 1013
 
Davies, R.L., Efstatiou, G., Fall, S.M., Illingworth, G., Schechter, P.L., 1983, ApJ, 266, 41
 
Davies, J.I., Phillipps, S., 1988, MNRAS, 233, 553
 
Davies, J.I., Phillipps, S., Cawson, M.G.M., Disney, M.J., Kibblewhite, E.J., 1988, MNRAS, 232, 239
 
Faber, S.M., 1973, ApJ, 179, 423

Faber, S.M, Lin, D.M.C., 1983, ApJ, 266, L17
 
Ferguson, H., Sandage, A., 1988, AJ, 96, 1520 (FS88)
 
Ferguson, H., Binggeli, B., 1994, Astron.~Astrophys.~Rev., 6, 67 (FB94)
 
Graham, A., Lauer, T.R., Colless, M., Postman, M., 1996, ApJ, 465, 534 
 
Hjorth, J., Madsen, J., 1995, ApJ, 445, 55
 
Ichikawa, S.I., 1989, AJ, 97, 1600
 
Jerjen, H., Tammann, G.A., 1996, A\&A, in press
 
Jerjen, H., Freeman, K.C., Binggeli, B., Beaulieu, S., 1996, in preparation
 
Kormendy, J., 1985, ApJ, 295, 73
 
 
Kormendy, J., Bender, R., 1994, in: {\em ESO/OHP Workshop on Dwarf Galaxies}, 
Meylan, G. \& Prugniel, P., eds., ESO, Garching, p.~161

Kormendy, J., Richstone, D, 1995, ARA\&A, 33, 581
 
Kormendy, J., Dressler, A., Byun, Y.-I., Faber, S.M., Grillmair, C., 
Lauer, T.R., Richstone, D., Tremaine, S., 1994, in: {\em ESO/OHP Workshop on Dwarf Galaxies},
 Meylan, G. \& Prugniel, P., eds., ESO, Garching, p.~147
 
Nieto, J.-L., Bender, R., Surma, P., 1991, A\&A, 244, L37
 
Peletier, R.F., 1993, A\&A, 271, 51
 
Ryden, B.S., Terndrup, D.M,, 1994, ApJ, 425, 43
 
Ryden, B.S., Terndrup, D.M., Pogge, R., Lauer, T., 1997, these proceedings
 
Sandage, A., Binggeli, B., 1984, AJ, 89, 919
 
Sandage, A., Fomalont, E., 1993, ApJ, 407, 14

Sandage, A. Tammann, G.A., 1995, ApJ, 446, 1
 
Sandage, A., Binggeli, B., Tammann, G.A., 1985, AJ, 90, 1759
 
S\'ersic, J.L., 1968, Atlas de galaxias australes, Observatorio Astronomico, Cordoba
 
Stein, P., Jerjen, H., Federspiel, M., A\&A, submitted
 
Tremblay, B., Merritt, D., 1996, AJ, 111, 2243
 
Vader, J.P., Sandage, A., 1991, AJ, 379, L1
 
Wirth, A., Gallagher, J.S., 1984, ApJ, 282, 85
 
Young, C.K., Currie, M.J., 1994, MNRAS, 268, L11
 
Young, C.K., Currie, M.J., 1995, MNRAS, 273, 1141 
 
\bye